# The excitation energy spectrum for a system with electron pairs tunneling in a two-leg ladder has a doping depended gap

Valentin Voroshilov

Physics Department, Boston University, Boston, MA, 02215, USA

A new model with a new Hamiltonian and a new canonical transformation is offered as the means for studying properties of a system of strongly correlated electrons. Consideration of the simplest possible situation, namely a system on non-interacting electrons in a two-leg ladder, leads to an expression for the excitation energy spectrum with no energy gap at the half-filling and with an energy gap away from the half filling.

74.20.Mn    71.10.Li

Since the time of the first high temperature superconductor was discovered[1], there is no yet a commonly accepted explanation of this phenomenon. Many publications on the matter start from some plausible reasoning leading to establishing of the model Hamiltonian and a discussion of the structure of the ground state. That plausible reasoning represents the physical view of the authors and, as long as the Hamiltonian and the ground state are set, the next step is using various mathematical methods to analyze the properties of the model. Many approaches are based on the Hubbard model[2]. The reason for using the Hubbard model is the fact that the parent state of a HTSC is an antiferromagnetic, which, when doped, exhibits many peculiar properties, including HTSC. However, the search for new models[3] is continuing and might lead to new insights on the matter and help to advance understanding of the nature of HTSC.

The author firmly believes that for every complicated physical phenomenon a clear and "simple" model exists which grasps the essence of the phenomenon.

For example, the model Albert Einstein offered to explain the photoelectric effect is very simple - from the mathematical point of view. Two Einstein's postulates of the theory of special relativity are also very simple – as long as one accepts the new view on space and time. Even the idea behind the Einstein's theory of General relativity becomes clear if one accepts the notion that time and space can bend: the more energy is concentrated the more space and time are bent. The Bohr's model of a hydrogen atom involves only elementary mathematics, but explained linear spectra. BCS theory of conventional superconductivity is based on a "simple" idea that electrons can form bound pairs.

In this paper, we offer a novel notion which leads to a "simple" model for understanding HTSC.

The model is based on the view that doping plays more important role than an electron – electron interaction (direct or mediated by some agent).

We start from a very well-known notion that in a single Hydrogen molecule, for two electrons with anti-parallel spins the wave-function has a solution with both electrons occupying the same location. From a formal point of view, it means that there are "instances" (i.e. tiny time intervals) when the electrons occupy "the same location" (i.e. very close to each other).

The similar statement can be done for electrons in the Cu-O bond in a cuprate-based HTSC. This pair of electrons can be seen as a bonded pair; but the paring happens purely due to the quantum properties of matter, without any specific mediating agent. We take this notion as a starting point for the further development of our view.

At the half filling, the charge density inside the material has the symmetry imposed by the symmetry of the lattice. Essentially, all locations "look alike". Let us assume that the number of electrons becomes less than the number of sites (this assumption does not affect further modelling). This leads to a formation of a local zone with deficiency of electrons. A zone of this kind becomes a zone of attraction for electrons around. However, in order to reach that zone, electrons have to overcome a potential barrier. Two electrons occupying neighboring sites and having opposite spins (due to the property of the parent material) might find themselves "momentarily" close to each other (which would not be possible for electrons with parallel spins) and become a "spin-zero-boson" which – in turn – can tunnel into the zone with the deficiency of electrons.

One should assume that (due to the structure of the material, including the difference in the spin-structure) the probability for a single electron to tunnel is less than the probability to tunnel for the pair.

This type of tunneling is not restricted to low temperatures, hence might be happening even above the critical temperature of HTSC. The conclusion on the absence or presence of a superconductive phase has to be done based on the analysis of the excitation spectrum together with the behavior of anomalous correlation functions.

Based on the presented view, one might assume that the ground state of the system should have the structure similar to the well know structure of the BCS[4] ground state, however paired electrons should not have opposite momenta (like in Cooper pairs) but instead, since they "travel" (tunnel) together (in the same direction), should have *the same momentum*.

Two mental pictures could help us to visualize the bonding process between the electrons, and to arrive at the Hamiltonian for the system. First, we can imagine two coupled gears rotating in opposite directions. The parts of the gears which are touching each other move in the same direction, i.e. have the same momentum, like the electrons assumed to be bonded in a HTSC (two electrons with opposite spin, opposite orbital momentum, but the same linear momentum, and located "close" to each other). Second, if we imagine a diatomic gas under such conditions that some of the molecules would be dissolved into individual atoms, the Hamiltonian of this gas could be written as a composition of the Hamiltonian for the subsystem of diatomic molecules, the Hamiltonian for the subsystem of individual atoms, and the interaction terms. This view will be used below to write the Hamiltonian for the electrons in a HTSC.

Let us start from thinking of the Schrodinger equation for *Ne* electrons. To make a transition to a second quantization one has to select a set of one-electron wave functions as the means for constructing Slater determinant. However, in anticipation of the existence of pairs of bonded electrons one could construct determinant using *Ne* – 2 one-electron wave functions and one wave function describing a bonded pair.

In this case, the resulting Hamiltonian would have kinetic energy term related to the motion of individual electrons, but also a kinetic energy term related to the motion of bonded electron pairs.

In this paper, the Hamiltonian in Eq.1 is restricted to the simplest possible case of non-interacting electrons in a two-leg ladder. The importance of the antiferromagnetic order is preserved in the structure of the term describing tunneling electron pairs.

The Hamiltonian neglects electron motion between the two chains, only the motion along each chain provides an input into the kinetic energy of the system.

In Eq.1, sites of a 2x*N* lattice are numerated with $k = 1, \ldots, N$ (in *x* – direction), and $n = 1, 2$ (in *y* – direction); $\sigma = \pm$ indicates the direction of the *z*-component of the electron spin; units are set with lattice constant $a = 1$, Boltzmann constant $k_B = 1$, and Planks constant $\hbar = 1$.

$$H = -t \sum_{k n \sigma} (a^+_{k+1\,n\,\sigma} a_{k\,n\,\sigma} + H.C) - \\ - v \sum_{k \sigma} (a^+_{k+1\,1\,\sigma} a^+_{k+1\,2\,-\sigma} a_{k\,2\,-\sigma} a_{k\,1\,\sigma} + H.C) - \mu \sum_{k n \sigma} a^+_{k\,n\,\sigma} a_{k\,n\,\sigma}. \quad (1)$$

In Eq.1 *t* is the hopping integral, $v$ is the analog of the hopping integral for tunneling electron pairs, $\mu$ is chemical potential (the last term is to remove the restriction on the number of electrons in the system), and *a*-operators are creation and annihilation operators for the electrons in the lattice. Hamiltonian in Eq.1 has the structure very similar to the structure of the Hubbard model. This might be the reason for the Hubbard model to be able to describe certain features of HTSC. The similarity between the models also leads to a conclusion that the mathematical analysis of the presented model might be of the same level of elaborating as the Hubbard model (even with all the simplifications used to arrive at Eq.1). However, in order to just get the first impression of the viability of the model one can build on the offered above hypothesis about the ground state of the system. For example, using the ground state wave function one can calculate the expectation energy of the ground state for Hamiltonian (1).

Instead, we will use a different but an equivalent approach of defining new operators using a canonical transformation equivalent to the structure of the ground state wave function.

The first step is to make a transition into the momentum space using standard introduction of creation and annihilation operators (*b* – operators) acting in the momentum space, i.e. Eq.2.

$$a_{k\,n\,\sigma} = \frac{1}{\sqrt{N}} \sum_p b_{p\,n\,\sigma} e^{-ipk} \; ; \; p = \pm \frac{2\pi m}{N}; \; m = 0, 1, \ldots, \frac{N}{2} \tag{2}$$

The new canonical transformation has to combine creation and annihilation operators for electrons with opposite spins but the same momentum by defining new creation and annihilation operators (*c* – operators); the assumed property of the new operators is that when an annihilation *c* – operator acts on the ground state vector of the system the result is zero.

This transformation, which is an equivalent of a well-known Bogoliubov[5] canonical transformation, is described by Eq.3.

$$b_{p\,n+} = u_p c_{p\,n+} + w_p c^+_{p\,n-} \, , \quad b_{p\,n-} = u_p c_{p\,n-} - w_p c^+_{p\,n+} \, , \quad u_p^2 + w_p^2 = 1 \,. \tag{3}$$

Note, that in Eq. 3 both *b*-operators and *c*-operators related to the same momentum *p*.

From this place forward, the calculations become routine, since this approach has been known for decades and is described in numerous publications, including textbooks[6].

In short, when Hamiltonian (1) is written in terms of *c* – operators, terms with the structure of *ccn* (and H.C.) are exactly eliminated by setting a specific condition on the variables $u_p$ and $w_p$ (via an equation also involving excitation density $n_{pn\sigma}$); all other terms which are non–linear in terms of excitation density $n_{pn\sigma}$ are neglected due to an assumption that at low temperatures excitation density $n_{pn\sigma}$ is almost zero. Then the Hamiltonian takes a form of the one describing the system of non–interacting "particles", i.e. quasiparticles with a certain excitation energy spectrum, $\varepsilon(p)$. In particular, if $\varepsilon(p=0)=0$, the excitation energy spectrum has no energy gap, but otherwise the gap exists. If in addition to the existence of the energy gap the anomalous correlation functions for electrons are also not equal to zero, that is a strong indication of the existence of the superconductive phase.

For the model above for the expiation energy spectrum, $\varepsilon(p)$ calculations lead to Eq.4.

$$\varepsilon(p) = \frac{4v}{N}(2w_p^2 - 1) \sum_\xi \cos(p + \xi)\,[w_\xi^2 + (2w_\xi^2 - 1)n_\xi] \,. \tag{4}$$

Calculation also shows that $w_p^2 = <E_0|b^+_{p\,n\,\sigma} b_{p\,n\,\sigma}|E_0>$ is equal to the density of electrons (not quasiparticles) in the momentum space. Considering the simplest possible scenario, as the zeroth correction to the properties of the system, we can assume that all electrons (which are non-interacting in this model) occupy all momentum space below a certain momentum, $p_F$, so for $|p| > p_F$, $w_p^2 = 0$, and for $|p| < p_F$, $w_p^2 = 1$ (i.e. a standard step-function).

In that case, one finds that $p_F = \pi n_e/2$ ($n_e = N_e/(2N)$ is the electron density in a real space), and the energy spectrum (4) becomes $\varepsilon(p) = \frac{4v}{\pi}\left(1 - \sin\left(p + \frac{\pi}{2}n_e\right)\right)$, with $\varepsilon(0) = \frac{4v}{\pi}(1 -$ $\sin\left(\frac{\pi}{2}n_e\right)$. In this model, at the half filling when $n_e = 1$, $\varepsilon(0) = 0$, hence there is no gap. For small values of doping $x = n_e - 1$ we obtain an approximation, $\varepsilon(0) = \frac{\pi v}{2}x^2$, which means that doping in any directions should lead to development on the gap in the energy spectrum.

If we calculate anomalous correlation function $< E_0 | b^+_{p\,n+} b^+_{p\,n-} | E_0 > = w_p u_p$, condition $u_p^2 + w_p^2 = 1$ makes it to be equal to zero.

However, it is naturally to expect that the actual electron distribution is not described by a simple step-function; for example, due to electron interactions the distribution will be spread above and below momentum $p_F$. In that case in addition to the gap in the excitation energy spectrum the system also will have non-vanishing anomalous correlation functions. This understanding asserts the feasibility of the model as one of the prospective models for studying the properties of HTSC.

If this picture is correct, experiments with cold atoms will not be able to demonstrate HTSC. The search should be directed to explain what properties of HTSC make "pair-bonding" and "pair" tunneling in those materials different from other doped antiferromagnetics.

A two-fluid phenomenological model is based on the use of two densities: the normal one should be "standard" electron density described by a Fermi-liquid (and exhibits the same properties in superconductors of all types). But the "super-fluid" electron component should differ depending on the type of a super conductor: in a BCS-type superconductor the peak value of the distribution for the momentum of bonded electron pairs is ZERO; but according to the proposed model, in HTSC this value should *not* be equal zero. One might expect that experiments with mechanically moving HTSC will demonstrate that the pairs of bonded electrons have non-zero linear momentum.

---

[1] J.G. Bednorz and K.A. Muller, Z. Physik **B 64**, 189 (1986).

[2] P.W. Anderson, The theory of superconductivity in the high-Tc cuprates (Princeton University Press, Princeton, N.J., 1997), p.20, 133.

[3] Philip W. Anderson, "Do we Need (or Want) a Bosonic Glue to Pair Electrons in High Tc Superconductors?", http://arxiv-web3.library.cornell.edu/pdf/cond-mat/0609040.pdf.

V. Voroshilov, Physics C: Superconductivity, **Vol 470, No. 21,** p. 1962 (2010 (Nov)).

[4] J. Bardeen, L. N. Cooper, and J. R. Schrieffer, "Microscopic Theory of Superconductivity", Phys. Rev. **Vol. 106**, p. 162 (1957).

[5] J.G. Valatin, Comments on the theory of superconductivity, in: N.N. Bogolubov (Ed.), The Theory of Superconductivity, International Science Review Series, **Vol 4** (Taylor & Francis, US, 1968) pp. 118–132.

[6] N.N. Bogolubov, V.V. Tolmachev, D.V. Shirkov, A New Method in the Theory of Superconductivity (Consultants Bureau, New York, 1959, Chapter 2, Appendix II).

**Appendix: the origins of the idea**

When I was heading toward my MS degree in theoretical physics, my thesis advisor was Yuri Abelevich Nepomnyashchiy. His study was on the superfluidity in liquid Helium.

In 1988 he published a book "Superfluid Bose-liquid with strongly correlated paired condensate".

The fact is that at $T = 0$ K, the Bose-Einstein condensate comprises just a small percentage of the liquid, but the whole liquid is in a superfluid sate. In a low density (dilute gas, week interaction) approximation, one can use Belyaev technique (http://www.jetp.ac.ru/cgi-bin/index2/gf-view/en/extending-methods-of-quantum-field-theory-to-problems-in-low-temperature-physics-jetp-papers-by-v-m-galitskii-a-b-migdal-and-s-t-belyaev-in-1958), which leads to the rise of anomalous Green functions, $<a_p a_{-p}>$ or $<a_p^+ a_{-p}^+>$, which describe correlations between two atoms with opposite momenta, $p$ and $-p$ (very similar to BCS model of superconductivity; I used this technique to study properties of a non-equilibrium dilute Bose gas).

But the combination of the Bose-Einstein condensate with the paired condensate still would not cover the total amount of the superfluid Helium.

Yuri Abelevich Nepomnyashchiy was working on the idea that at $T = 0$ K, Helium can be understood as a combination of many condensates: a one particle condensate (the "standard" Bose-Einstein condensate composed from particles with $p = 0$); the paired condensate (composed from particles with opposite momenta); then the condensate composed from three correlated particles with total momentum equal to zero; then a four-particle condensate, etc., he called it a super-condensate. All condensates together, i.e. the super-condensate, produce the superfluid liquid. When temperature start rising, condensates are being gradually destroyed.

All existing condensates (i.e. existing super-condensate) compose the superfluid component of Helium, and the rest constitutes the "normal" component, in an agreement with the phenomenological two-fluid Landau model.

Naturally, he was not the first one who was exploring microscopic and phenomenological two- (or more) component models of superfluidity or superconductivity (e.g. R.P. Feynman, https://journals.aps.org/pr/abstract/10.1103/PhysRev.94.262; J. Bardeen, https://journals.aps.org/prl/abstract/10.1103/PhysRevLett.1.399).

In ALL those models "superior" component (responsible for superfluidity or superconductivity) is composed of individual or correlated particles with **zero total momentum.**

<div align="center">But does it *have* to be zero?</div>

What if instead, the "superior" component is composed of *macroscopic* streams which travel in *opposite* directions? The total momentum of the whole system still will be equal to zero.

Naturally, the properties of such system would be very different from the properties of a "standard" superfluid or superconductive matter.

**Could high temperature super conductors be such materials?**

I think, that is a question which is worth to explore.

I am not a physicist (although, I managed to publish in "Physica C; Superconductivity"), but a physics graduate, and the mystery of HTSC fascinates me, I just cannot not think about it, hence, this paper.

I know that saying that my mathematical apparatus is very limited would be an understatement; and that it is not nearly enough to further the analysis to measurable results. But I also know that the idea itself (a) has the same right to exist as only other ideas; (b) is at least peculiar enough to be worth to be worked out in more details.